\documentclass[submission]{eptcs}
\providecommand{\event}{Festschrift Symposium for Dave Schmidt 2013} 
\usepackage[T1]{fontenc}
\usepackage[utf8]{inputenc}
\usepackage{textcomp}
\usepackage{graphicx}
  \graphicspath{ {figures/} }
\usepackage{grffile}
\usepackage{flushend}
\usepackage{todonotes}
\usepackage{listings}
\usepackage{enumerate}
\usepackage{amsmath}
\pdfpageattr {/Group << /S /Transparency /I true /CS /DeviceRGB>>}

\newcommand{\code}{\texttt}
\newcommand{\newnotion}{\emph}

\newcommand{\origttfamily}{}
\let\origttfamily=\ttfamily
\renewcommand{\ttfamily}{\origttfamily \hyphenchar\font=`\-}

\begin{document}
\title{Higher-Order Process Modeling:\\
Product-Lining, Variability Modeling and Beyond}
\author{Johannes Neubauer \quad Bernhard Steffen \quad\qquad Tiziana Margaria
\institute{\hspace*{8em} Chair of Programming Systems 	  \hspace*{5em} Chair of Service and Software Engineering\\
	   \hspace*{4em}  Technische Universität Dortmund \hspace*{8em} Universität Potsdam\\
	   \hspace*{6em} Germany 			  \hspace*{16em} Germany\\
		\email{\{johannes.neubauer, steffen\}@cs.tu-dortmund.de} \email{\{margaria\}@cs.uni-potsdam.de}}
}
\def\titlerunning{Higher-Order Process Modeling}
\def\authorrunning{J. Neubauer, B. Steffen, T. Margaria}
\maketitle
\begin{abstract}
We present a graphical and dynamic framework for binding and execution of (business) process models. It is tailored to
integrate 1) ad hoc processes modeled graphically, 2) third party services discovered in the (Inter)net, and 3)
(dynamically) synthesized process chains that solve situation-specific tasks, with the synthesis taking place not only
at design time, but also at runtime. Key to our approach is the introduction of \emph{type-safe stacked second-order
execution contexts} that allow for higher-order process modeling. Tamed by our underlying strict service-oriented notion
of abstraction, this approach is tailored also to be used by application experts with little technical knowledge: users
can select, modify, construct and then pass (component) processes during process execution as if they were data. We
illustrate the impact and essence of our framework along a concrete, realistic (business) process modeling scenario: the
development of Springer's browser-based Online Conference Service (OCS). The most advanced feature of our new framework
allows one to combine online synthesis with the integration of the synthesized process into the running application.
This ability leads to a particularly flexible way of implementing self-adaption, and to a particularly concise and
powerful way of achieving variability not only at design time, but also at runtime.
%
\end{abstract}
\section{Motivation}
Business process modeling has evolved from mere collection of requirements artifacts, supported by tools like
ARIS~\cite{aris}, to an important element in the whole development process of complex applications. With the
introduction of execution semantics for BPMN2~\cite{bpmn2spec,allweyer2009bpmn}, business process models have started
to be used to integrate the application expert in the overall development process. This inclusive trend is however
still young and suffers from complicated mechanisms for integrating user-defined or third party
activities/services~\cite{bpm-service-integration,activiti,jBPM,apache-ode,oracle-bpel}. The only exception is the
jABC framework, that was designed from the very beginning for easing the integration process and that even offers
systematic formal methods-based product line ~\cite{Linden:2007:SPL:1296141} and
variability~\cite{JoeLMSS2012} support. On the other hand, in particular AristaFlow~\cite{AristaFlow} promoted ad hoc process modeling as a particularly convenient way for customization. An ad-hoc approach allows users to adapt processes at runtime for a single use, a heuristics that has also successfully been used in~\cite{Lampre2012} Sec.~3.2.3.

What is missing to be fully prepared for future demands is a comprehensive, type-safe framework for flexibility, able to
capture variability uniformly both at design time and at runtime in a hierarchical fashion. The usefulness of this ad-hoc and context-aware flexibility is illustrated by the
following three increasingly demanding route planning scenarios:
\paragraph{Static Route Planning:} Planning the route before a trip using e.g. google maps gives a selection of nice choices and the option for a detailed
investigation, and one is able to `personalize' the own trip. However, the choice of route is fixed as soon as it is
printed out to leave (of course one may have printed a couple of alternatives, one never knows...). Route planning in this style is very similar to classical variability modeling in software products: the flexibility strikes at design time. This mindset is widely popular, it is also espoused in offline planning and in system
synthesis. E.g., in HW/SW codesign the partitioning of what goes in hardware and what goes in software happens early in the design and is basically irrevocable. After this planning there is a dedicated, irrevocable step, e.g. making the printout, or, in BPM scenarios, compiling the model to a running system which then remains unchanged.
\paragraph{Dynamic Route Planning:}  Today's navigation systems flexibly adapt to changes at runtime (online planning).
E.g., once missed an exit, the navigation system will replan from there. True replanning from scratch with changed frame
conditions (blocked road, ad hoc changes of requirements) is optimal e.g. for fault tolerance.\footnote{The best
one can do with offline systems is to have fixed preplanned alternatives for certain foreseeable faults, which is essentially offline planning:  make a couple of printouts of different routes, or in the BP scenario precompute some alternatives to switch between). It is the dominant approach in BP practice, but merely covers foreseen
problems.} This dynamic approach is much more flexible, helps in any situation which can be
captured with the available resources, but requires true runtime modification and consequently a much faster, typically
automatic decision process for the best choice (unlike the offline google scenario, where one could calmly investigate
all options, like scenery, restaurants etc. and choose to one's liking).\footnote{Of course, Google meanwhile also supports
navigation on its mobile devices \ldots.} In this scenario, one should have a very good `default' for automatic
decision, because the manual interaction is limited at runtime: there are time constraints (driving and not
willing to stop), and there are resource constraints (the navigation system is not a full computer
equipment, or one has reduced permissions, hindering to start the full machinery). Thus this approach
also delivers process-level self-adaptation.
\paragraph{Route Planning with Higher-Order elements:} Future travel systems might combine a classical navigation system with various supporting systems that help one to control, steer, or react to specific situations. Traffic or weather forecast systems, roadside assistance, speed control services, electronic toll systems,  ferry operations etc.: many such systems are by their nature local, so future travel systems should
automatically adapt to the supporting systems currently offered to provide a comprehensive support. Ideally,
discovery services will detect locally offered supporting services, to be then dynamically integrated
as part of the travel system. This means that the future travel systems should be \emph{parameterizable with the supporting
systems} required to complete their functionality in a situation-specific fashion.

Thus future travel systems like this comprise functionality/sub-processes (e.g. service discovery) that provide new
concrete services/processes (e.g. a service for paying toll electronically) that are passed to other (sub-)processes (as a
parameter) for completing their tasks, e.g. to automatically use the provided payment service as part of the future travel system to pay the toll. In such a {\em second-order} structure, only services/processes can be passed around that are themselves first order (i.e. only have data parameter).

The approach we are proposing in the following is not limited to this second-order scenario, but
it also allows a controlled form of {\em higher-order} structures: think of the discovery service/process to be itself a
parameter of the future travel systems, which may be dynamically exchanged at need for a better or more specialized discovery service.
\subsubsection*{Contributions}
In this paper we present a graphical framework for dynamic binding and execution of (business) process models. It is
tailored to support the full flexibility described above by introducing \emph{type-safe stacked second-order execution
contexts} that allow higher-order process modeling.

Higher-order is here tamed by our underlying strict service-oriented notion of abstraction\footnote{Our notion of higher
order avoids many of the well known problems of full higher-order (cf. \cite{Langmaack73,Schmidt96}) as it is based on a
very strong notion of type correctness (cf. \cite{Schmidt94}).}, which supports a localized form of type checking. This
approach is tailored also to be used by application experts to profit of the smooth integration of
\begin{itemize}
\item ad hoc processes - modeled graphically,
\item third party services discovered in the (Inter) net, and
\item (dynamically) synthesized process chains solving situation-specific tasks.
\end{itemize}
Users can select, modify, construct and then pass (component) processes during process execution as if they were data.

Our target language is Java. However, Java is only used here at the `tool/system-level', e.g. to support
consistent typing and other object-oriented features like polymorphism. The implementation language for the individual
processes may be indeed a different language. All the presented concepts are quite general and they could
have been implemented in and for other languages like \newnotion{C++} or \newnotion{C\#}.\medskip

\noindent
We will illustrate the impact and essence of our framework along a concrete, realistic (business) process modeling scenario: the development of Springer's browser-based Online Conference Service (OCS), a complex model-driven
and service-oriented manuscript submission and review system. The OCS deals with hundreds of conferences and thousands
of users of quite different scientific communities, with quite different expectations. Thus variability was here a major design concern. We address it by:
\begin{itemize}
\item providing separate process models for specific tasks and procedures in the system, which can be
combined into larger processes,
\item allowing to flexibly include third party services, and
\item supporting ad hoc process modification of running processes by the conference organizers.
\end{itemize}
The most advanced feature our new framework allows one to combine online synthesis with the integration of the synthesized process into the running application (cf. Sec.~\ref{sec:ho-synthesis}). This ability leads to a particularly
flexible way of  implementing self-adaption, and to a particularly concise and powerful way of achieving
variability not only at design time, but also at runtime.\medskip

\noindent
{\bf Dedication:}
People recognize each other by their gait. In case of David Schmidt it was the common 
quest for the essence, or as we now often coin it, the quest for {\em simplicity}. One
major goal here is the move from the {\em How} to the {\em What}, as e.g.,
inherent in the step from operational to denotational semantics or, and this 
connected us most,
from data-flow equations, which characterize a data-flow problem indirectly
via fixpoint computation, to specifications in temporal logic, which directly 
specify the desired property. {\em Service-orientation} and {\em process modeling}
introduce another kind of {\em What} level with new opportunities and challenges, many of
which would strongly profit from Dave's broad experience, expertise and unbiased attitude, 
and we would like to invite him to join us in this exciting adventure.
\medskip

\noindent
In the following, Sec.~\ref{sec:prelim} will introduce our running example (the OCS), its variant rich
processes, and our graphical binding and execution framework for (business) process models, the jABC.
Sec.~\ref{sec:relwork} will discuss related modeling approaches, and  Sec.~\ref{sec:second-order}
will sketch two sides of second-order process modeling, that focus on activity/service passing and (sub-)process
passing respectively. Subsequently, Sec.~\ref{sec:higher-order} will present the tamed form of
higher-order process modeling that arises from introducing stacks of second-order contexts, and illustrate its
power along our running example. Finally, Sec.~\ref{sec:concl} will present our conclusions and perspectives.

\section{Preliminaries}
\label{sec:prelim}
We show the power of higher-order process modeling in the design of complex dynamic applications using a simplified 
situation of modifying the OCS, a complex web application for conference submission management, and the jABC as 
its design framework.
\subsection{The OCS}
\label{sec:ocs}
\begin{figure}[t]
\centering
\includegraphics[width=0.40\textwidth]{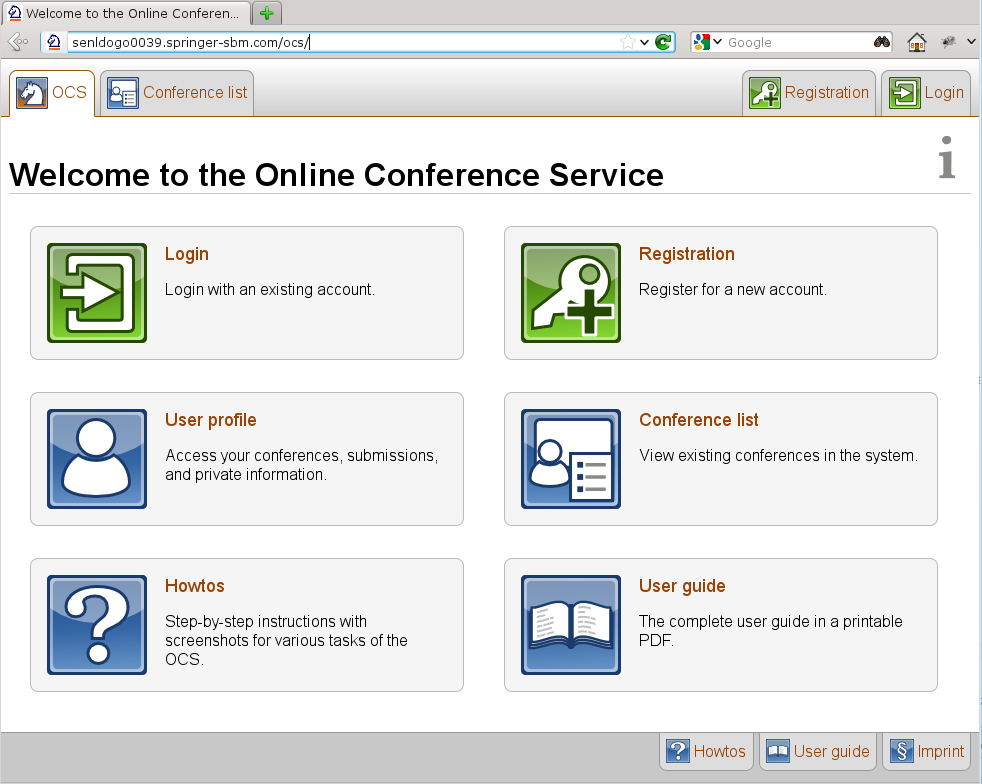}
\caption{Welcome page of the current OCS.}
\label{fig:ocs-welcome}
\end{figure}

The Online Conference Service (OCS) is an online manuscript submission and review service that is part of a long-lived software product line for the Springer Verlag\footnote{http://www.springer.com/computer/lncs?SGWID=0-164-6-447109-0}. Initially started in 1999~\cite{DBLP:journals/entcs/KarusseitM06}, the product line evolved over time to include also journal and volume production preparation services. In 2009 we started a radical redevelopment (see Fig.~\ref{fig:ocs-welcome}) to produce a new system designed for verifiability, as described in more detail
in~\cite{FMICSHBOCS11}.

The service needs to appropriately handle a wealth of independent but often indirectly related user interactions, where
accomplishing a task may have consequences for other participants or shared objects. From this point of view the OCS is
a user and context driven reactive system with a web interface. Users can decide whether they reject a task, which
typically consists of performing small workflows, when to execute it, and in case of multiple tasks in which order to process
them. The OCS is a role-based and context-sensitive system~\cite{karusseit2007web}: in a given state different roles may
have distinct action potentials. The large number of involved participants, roles (like author or PC chair), interaction
styles, and the high degree of freedom in choosing individual tasks makes the OCS a challenge for business process
modeling approaches. The readers of this paper are scientists themselves, thus familiar with this kind of services and
the underlying workflows. We are ourselves both developers and users of the system -- and therefore application experts
as well.

\begin{figure}[t]
\centering
\includegraphics[width=\textwidth]{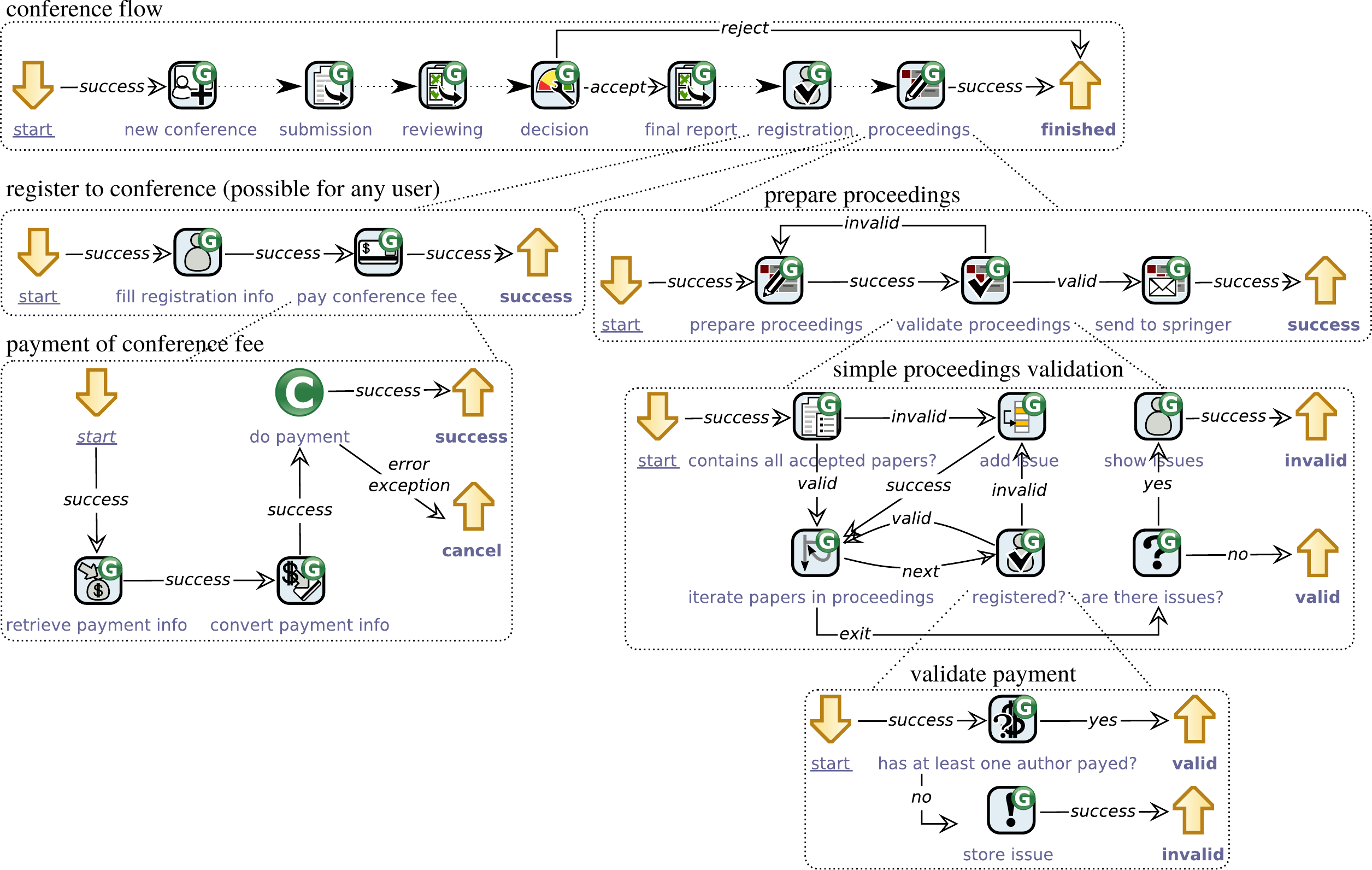}
\caption{An abbreviated conference flow of the OCS with sub-processes for the `registration' and `proceedings
preparation'.}
\label{fig:static-ocs}
\end{figure}

The new OCS is in productive use since 2009. The application supports the workflows for planning physical conferences
collaboratively in the web in so called \newnotion{conference services} (see `conference flow' in
Fig.~\ref{fig:static-ocs}), that proceed by and large as follows:
\begin{itemize}
 \item an admin sets up a new conference service
 \item authors submit online their papers to the conference,
 \item a PC chair (the head of the conference) assigns reviewers to the papers,
 \item reviewers access the assigned papers and submit their reports with an evaluation,
 \item the PC chair decides in coordination with the program committee members and the reviewers whether each paper should
  be accepted or rejected
 \item a final report containing the decision outcome is sent to the corresponding authors by the PC chair,
 \item authors of accepted papers upload a final version (also known as `camera ready') to the application.
\end{itemize}
The workflows adopted by various communities for organizing a conference in a service
like the OCS can differ considerably. Some have a dedicated `abstract submission' phase, some a combined `submission and
upload' phase. Some have a `bidding' phase, others directly assign reviewers to papers. Some use several review
cycles with a rebuttal phase for the authors. Journal services are in principle similar, but have no global phases and are organized along paper-centric workflows. Furthermore there are some more subtle differences with broad consequences, like normal, blind or double blind reviewing, or conferences structured with tracks and sessions. All these kinds of services run on the same system and
interchange some information, like e.g. the user accounts, which are global to the entire OCS.

These requirements already induce a large variability space, but there is more: the system has been enhanced with a
\newnotion{proceedings' production service} (PPS) for the Springer Verlag, where PC chairs can
\begin{itemize}
 \item add general information of a proceedings volume like title, volume number and preface,
 \item create topical parts for structuring the submissions,
 \item add papers of the conference to the topical parts and sort them,
 \item create arbitrary committee tables and add members to them,
 \item generate a preview PDF containing the general information, a table of contents, the committee lists, papers, and
  an author index, and
 \item download an archive of the complete proceedings prepared in conformity to the Springer Verlag's guidelines for
post processing and print.
\end{itemize}
The PPS mandates that the OCS integrate into the bigger context of the Springer production ecosystem: it must
interoperate with other Springer applications and third-party services in the web. In particular, the proceedings'
archive should directly be uploaded to a Springer Verlag FTP server (see node `send to springer' in the process model
`prepare proceedings' of Fig.~\ref{fig:static-ocs}), this way easing the interaction between the PC chair and the volume
responsibles. A proceedings manuscript should also conform to several constraints (see process model `simple proceedings
validation' in Fig.~\ref{fig:static-ocs}), e.g. that:
\begin{itemize}
 \item all accepted papers are contained in a topical part and
 \item for every paper in the proceedings at least one author has registered to the conference.
\end{itemize}
The latter is realized by testing whether at least an author has payed the conference fee (see process model `validate
payment' in Fig.~\ref{fig:static-ocs}) and therefore he or she is expected to attend and present at the conference.

In Fig~\ref{fig:static-ocs} the registration happens in the `register to conference' subprocess model of the `conference flow'. Here
users fill in the registration info and pay the conference fee. For better readability we omitted the \emph{1 to n}
relation between these process models and the exception handling. The sub-process model `payment of conference fee'
retrieves the payment information of the user, converts it to the data types of a specific external payment service, and
calls the external service.

The need for variability is apparent also here: each user might set a default
payment service, denoted by \newnotion{configuration variability}, or choose dynamically a payment service, called
\newnotion{manual selection variability}. This could be realized via a big if-clause or switch-clause in the model, as
is frequent in operating systems, in explicit product configuration management of product lines, and in conditional
programming. But if further payment services need to be integrated, with these hardcoded options the model must be adapted and redeployed
to the server. This becomes an issue in particular for long-running processes: the change might be then available only for
new conference services (after deployment) or running process models might need a possibly complex and risky on-the-fly
migration procedure. We will show in Sec.~\ref{sec:2ndorder-pm} how this can be realized like a charm via second-order
process modeling.

During the productive operation of the OCS more and more requirements arose for the proceedings preparation and the
properties a valid proceedings has to fulfill. This knowledge is mission critical, since mistakes in an already
printed proceedings are expensive and printing lies at the very heart of the business for our project partners at the Springer Verlag.
Hence, an increasing amount of validations like
\begin{itemize}
 \item are all necessary artifacts available (e.g. final version, copyright form, source archive)?,
 \item do the submitted sources compile?,
 \item does the final document adhere to the margins of the directed style?, and
 \item is the submission a plagiarism?
\end{itemize}
that depend on the community and conference settings should be executed on the prepared proceedings. Furthermore
some validations regarding the author registration can only happen if the used external registration service offers
them. E.g., to detect early potential `no shows', some registration services support to check whether a registered
author has also booked a flight and a hotel. We propose an elegant way to address these problems with higher-order
process modeling in Sec.~\ref{sec:higher-order}.
\subsection{XMDD in the jABC}
\label{sec:jabc}
The executable process models of the OCS are realized in the \newnotion{jABC}~\cite{Agile-IT,MarSte2004,StMaNa2006}, a framework
for service-oriented design and development. We describe our models as graphs since they have a graphical representation
as directed graphs, as this often is easier to comprehend and use for people without technical background. We will still
use the term `model' or `process model' if we want to emphasize that we are talking about executable business process
models. The jABC framework follows the \newnotion{extreme model driven design} (\newnotion{XMDD}) of~\cite{MargariaS12}.
XMDD combines ideas from service orientation, model driven design and extreme programming and enables application experts
to control the design and evolution of processes during their whole life-cycle on the basis of lightweight process
coordination (LPC)~\cite{MarSte2004}. The jABC allows users to develop services and applications easily by composing
reusable building blocks into (flow-) graph structures that are both formally well-defined and easy to read and build.
These building blocks are called Service Independent Building blocks (SIBs) in analogy to the original telecommunication
terminology~\cite{HLBand}, and in the spirit of the service-oriented computing paradigm~\cite{ServiceBeholder} and of
the \newnotion{one thing approach}~\cite{OTA-BPM09}, an evolution of the model-based lightweight coordination approach
of~\cite{MarSte2004} specifically applied to services. The one thing approach provides the conceptual modeling
infrastructure (one thing for all) that enables all the stakeholders (application experts, designer, component experts,
implementer, quality insurers, \ldots) to closely cooperate following the extreme model driven design paradigm. In
particular it enables immediate \newnotion{user experience} and \newnotion{seamless acceptance}, which is a central constituent of the one thing approach: The fact that all stakeholders work on and modify one and the same thing allows
every stakeholder to observe the progress of the development and the implications of decisions at their own level of expertise.

On the basis of a large library of such SIBs, which come in domain-specific collections as well as in domain-independent collections, the user builds models for the desired system in terms of hierarchical
service logic graphs (SLGs)~\cite{HierarchicalSD-97}. These graphs form the modeling backbone of the one thing approach.
All the information concerning documentation, role, rights, consistency conditions, animation code, execution code,
\ldots, come here together. Immediate user experience is a result of the extreme model driven design approach, where
already the first graphical models are executable, be it as the basis for interactive `what/if games', documentation
browsing or simple interpreted animation. This allows one to detect conceptual errors early in the requirement models.

SLGs are also directly formal models: they are semantically interpreted as Kripke Transition Systems (KTS), a generalization of both Kripke structures and
both Kripke structures (KS) and labeled transition systems~\cite{MuScSt1999} (KTS) that allows labels both on nodes and 
edges. A KTS over a finite set of atomic propositions $AP$ is a structure $M = (S, Act, T, I)$, where
\begin{itemize}
\item $S$ is a finite set of \textit{states}.
\item $Act$ is a finite set of \textit{actions}.
\item $T \subseteq S\times Act\times S$ is a total \textit{transition relation}.
\item $I : S \to 2^{AP}$ is an \textit{interpretation function}.
\end{itemize}
A KS is a KTS with an empty set of actions, an a LTS is a KTS with a trivial interpretation $I$.

Nodes in the SLG represent
activities (or services/components, depending on the application domain). The edges directly correspond to SIB branches:
they describe how to continue the execution depending on the result of the previous activity. The SIBs communicate via a
shared memory concept. Its incarnation is called \newnotion{context}, it is itself hierarchical and addresses data
objects with identifiers denoted by~\newnotion{context variables}. In an SLG, a SIB may represent a single functionality
or a whole subgraph (i.e., another SLG), thus serving as a macro that hides more detailed process models. SIBs are
parameterizable, so their behavior can be adapted depending on the current context of use.

This approach grants a high reusability not only of components but also of whole (sub-)models within larger
applications. Furthermore it encourages separation of concerns and supports divide and conquer as an aid, in order to
come to a tractable problem solution.

\begin{figure}
 \centering
  \includegraphics[width=.5\textwidth]{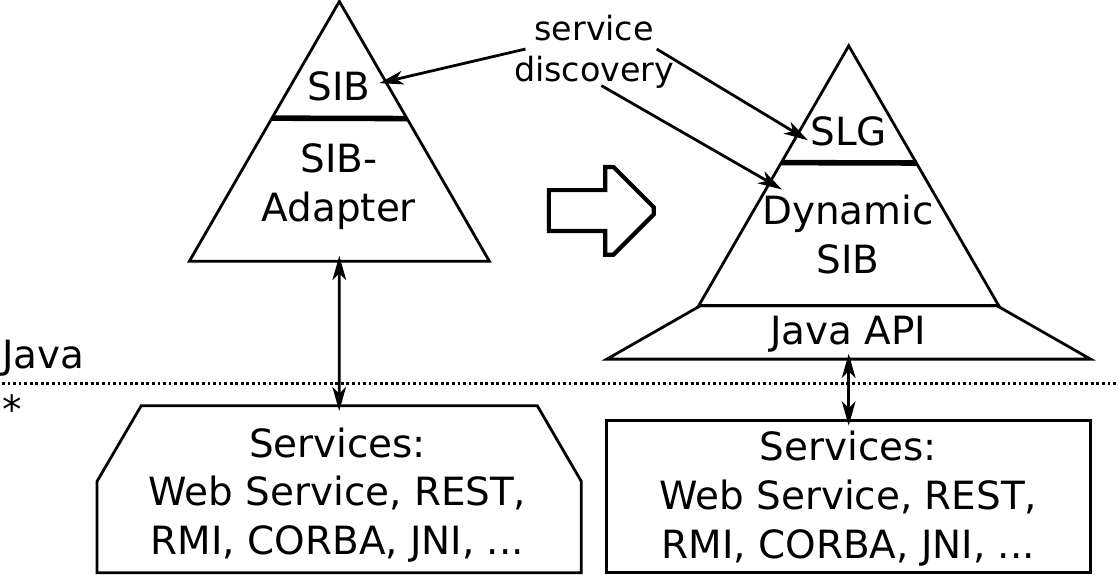}
  \caption{Reorientation from SIB adapter (left)  to dynamic SIB pattern (right).}
  \label{fig:adapter-dynamic-pattern}
\end{figure}

Technically, domain-specific business activities that are made available as ready to use services or as software components are provided within the jABC as SIBs via an adapter
pattern~\cite{StMaNa2006} (see left triangle in Fig.~\ref{fig:adapter-dynamic-pattern}).
An arbitrary service, e.g. an internal or external service, is integrated at the SIB level by a technical expert who implements two
artifacts:
\begin{enumerate}
  \item a new \newnotion{service independent building block} (\newnotion{SIB}) in form of a Java class defining
  the parameters, documentation, and appearance of the activity and
  \item a SIB adapter in form of a static method for processing the parameters. The adapter calls the
  service (or a bunch of services, for adapters that are more abstract than a single service invocation), and evaluates the
  result in order to determine the successor activity.
\end{enumerate}
The adapter pattern enforces the SIB-level compatibility of otherwise technically heterogeneous, even arbitrary domain-specific activities which interact via their common context of resources.

These business activities are organized in taxonomies, so they can be easily discovered by algorithms that search for
activities that satisfy a certain profile (given as a logic formula ranging over the taxonomies) and then (re)used for building the SLGs of complex business process models in the corresponding graphical development environment.
\begin{figure}
\centering
\includegraphics[width=.8\textwidth]{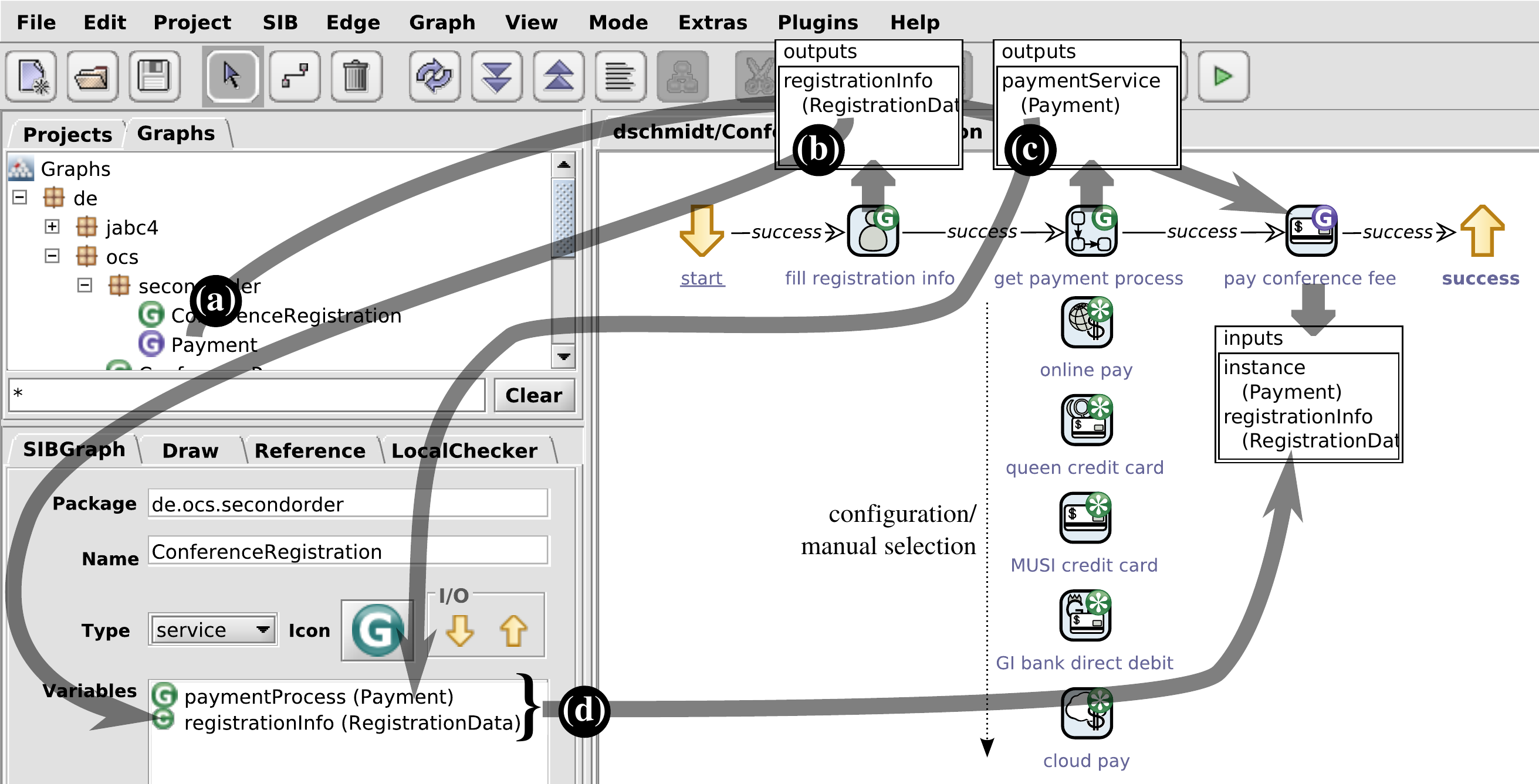}
\caption{The main window of the jABC framework. The grey arrows
indicate how second-order process modeling helps to deal with the
variability of the `register to conference' process model of Fig.~\ref{fig:static-ocs} regarding the payment of the
conference fee (cf. Sec.~\ref{sec:2ndorder-pm}).}
\label{fig:2ndorder-payment}
\end{figure}

Fig.~\ref{fig:2ndorder-payment} shows how the second-order payment
process model described in Sec.~\ref{sec:2ndorder-pm} is presented to a jABC user:
\begin{enumerate}
 \item \newnotion{browser area}(top left): a tabbed pane contains a \newnotion{project browser} and a \newnotion{sib browser}.
These components offer resource trees which enable to browse e.g. for SLGs and SIBs.
 \item \newnotion{inspector area} (bottom left): a tabbed pane contains several \newnotion{inspectors} that give
context information to the currently selected components like nodes in a graph or the graph itself. Some
inspectors support editing the information, too. Currently the \newnotion{SIBGraph inspector} is open and shows
information related to the selected SLG.
 \item \newnotion{graph canvas} (right): this is the modeling area. It displays the graph structure of the current SLG and offers
functionality to manipulate it.
\end{enumerate}
\section{Related Modeling Approaches}
\label{sec:relwork}
Service discovery has today a significant impact in any context aware platform. Finding out what concrete service
offerings are there in a certain location or neighborhood (by near field communication, bluetooth, or other
technologies) has become a standard capability offered by or on top of operating systems. For instance, the Google APIs
Discovery Service~\cite{googlediscovery} helps programmers to build client libraries, IDE plugins
and other tools that interact with these APIs. It provides a lightweight, JSON-based API that exposes machine-readable
metadata about Google APIs including machine-readable ``Discovery document'' for each of the supported APIs. This is an
excellent technical basis for the creation of SIBs, which in many cases can be even automated. We have in fact created a
number of SIB importers from Web services, e.g. used in the Semantic Web Service
Challenge~\cite{Cubczak2013,petrie2009semantic,swscWebsite}, or from the EMBOSS bioinformatics tool
suite~\cite{LaNaSM2009}, or for
SAP's R3 proprietary BAPI~\cite{DBLP:conf/sew/DoedtS11}. These services however are then manually annotated for
classification in the ontologies or taxonomies used as knowledge basis for the profile description and matchmaking by
the automatic service composition tools, like PROPHETS~\cite{NaLaSt2012}, the selection of \cite{MaMKIS2009}, or our
older approaches~\cite{MarSte2007,FrMaSt1994,StMaBr1997}, or other planning based approaches, like
e.g.~\cite{SiPWHN2004}.

Especially in the service oriented community, service discovery, matchmaking, and binding have been studied for years.
Discovery has been addressed by means of profile-based characterization followed by matchmaking between profiles of
service requests and service offers~\cite{swsmiAamics,swsWSMOLX,swsDIANE}. The kind of information offered in the
profile is here the discriminant for the quality of discovery. We used in~\cite{KuMaWi2007} discrete characterizations
of services, that went well beyond the pure syntactic compatibility of interfaces at the programming language level,
while others allowed also continuous or fuzzy characterizations~\cite{KueKoe2007}. The issue is then, what part of the
profile the matchmaking algorithm can consider and evaluate.

The annual International Contest on Semantic Service Selection (S3)~\cite{S3} monitors the state of the art advancement
in retrieval performance of service matchmakers for semantic web services. Recall, precision, response time are measured
over given test collections for prominent semantic service formats such as OWL-S, WSML and the standard SA-WSDL. Service
collections and service directories are also starting to appear, e.g., the OPOSSUM Portal~\cite{KueKoeKr2008}. However,
there is currently a lack of semantically annotated implemented services: since discovery and matchmaking only need the
service description, most service descriptions in the semantic matchmaking benchmark sets do not have a corresponding
implementation, making it impossible to really bind and execute the selected services. Approaches for automatic
annotations of real implemented and published services are starting to emerge~\cite{seekda}, but mostly the annotation
work is still manual and knowledge-intensive.

The approach described in this paper generalizes the typical implementations and approaches for flexible modeling of
hierarchical business process models:
\begin{itemize}
  \item Aristaflow~\cite{AristaFlow,springerlink:10.1007/978-3-642-12186-9-50} allows for hierarchical process
  definitions, declaration of input and  output behavior of a process (called graph template), and has a kind of local
  context that is type-safe for dealing with data objects in terms of data bases, i.e., data structures similar to
  \emph{struct} in the programming language C or simple records. All other types are hidden beneath
  \code{java.lang.Object}. The data-flow is directly modeled in the graph template. However every sub-process has only
  one result branch and therefore only one set of return values. Branching is done via XOR-nodes and other dedicated
  control-structure nodes conceptually similar to gateways in BPMN. Thus the Aristaflow is entirely first-order and
does not support polymorphism.
  \item Engines following the BPMN specification~\cite{bpmn2spec} like Activiti~\cite{activiti} and jBPM~\cite{jBPM}
  bind sub-processes statically to activities and therefore do not support second- or higher-order process modeling.
  In advance second-order process modeling on the service-level is not supported. Admittedly these solutions allow
  for execution of methods on objects in the shred resources. But, in contrast to our approach this is realized as a
  kind of script activity, i.e., the method call is written as an expression. Input parameters and return value are
  defined in the expression. Hence, the model is neither aware of the type of the service instance, its input parameters
  or the return values nor of which context variables are accessed.
\end{itemize}
\section{Second-Order Modeling}
\label{sec:second-order}
In this section we illustrate the effect of introducing second-order contexts in our modeling framework jABC.
Using service instances (in form of Java objects) just like data in the shared resources (context)
of a process instance allows business activities to execute virtual methods on them, which essentially
leads to a notion of {\em virtual services}. Going one step further by allowing whole process instances
in second-order contexts analogously introduces true second-order process modeling. We will see
that this elegantly allows one to enable variability, e.g. for the registration process in Fig.~\ref{fig:static-ocs}.
\subsection{Second-Order Services}
\label{sec:2ndorder-services}
To achieve service-level second-order process modeling we added support for dynamic binding of polymorphic,
domain-specific business activities~\cite{DBLP:conf/icsob/NeubauerS13,NNLSJM2013} (see right triangle in
Fig.~\ref{fig:adapter-dynamic-pattern}) in terms of dynamic service independent building blocks: \newnotion{dynamic
SIBs} or \newnotion{virtual SIBs}. These activities bind directly to a method e.g. of a class, interface or enumeration. 
They are organized in \newnotion{service trees} in the browser area of the jABC, a graphical tree structure respecting 
the package structure of the underlying Java classes. They can be easily discovered and (re)used for building complex 
business process models, the next generation service logic graphs (SLGs). This approach extends the adapter concept, 
where the service execution had been encapsulated in a static adapter, to a discipline where arbitrary service instances 
are part of the shared resources and can be used as input and output parameter just like data. It thus introduces the 
concept of {\em virtual activities} (represented by virtual/dynamic SIBs).

Since in Java all but primitive data types are objects, even domain-specific data objects like a \code{User} or
\code{Paper} may be used to execute services on it as `get' and `set' methods. These service-level business
activities are used to model \emph{technical SLGs}, which can be integrated in application level SLGs via business activities that
abstract from the service execution just like the adapter SIBs did before. Of course, these technical SLGs are not
limited to simple wrappers of a single service execution handling the parameter transfer. They may also transform input and
output parameters for executing external services, discover a concrete external service at runtime or aggregate the
execution of several services in order to introduce more sophisticated abstraction already at the technical level.

\begin{figure}[t]
\centering
\includegraphics[width=0.6\textwidth]{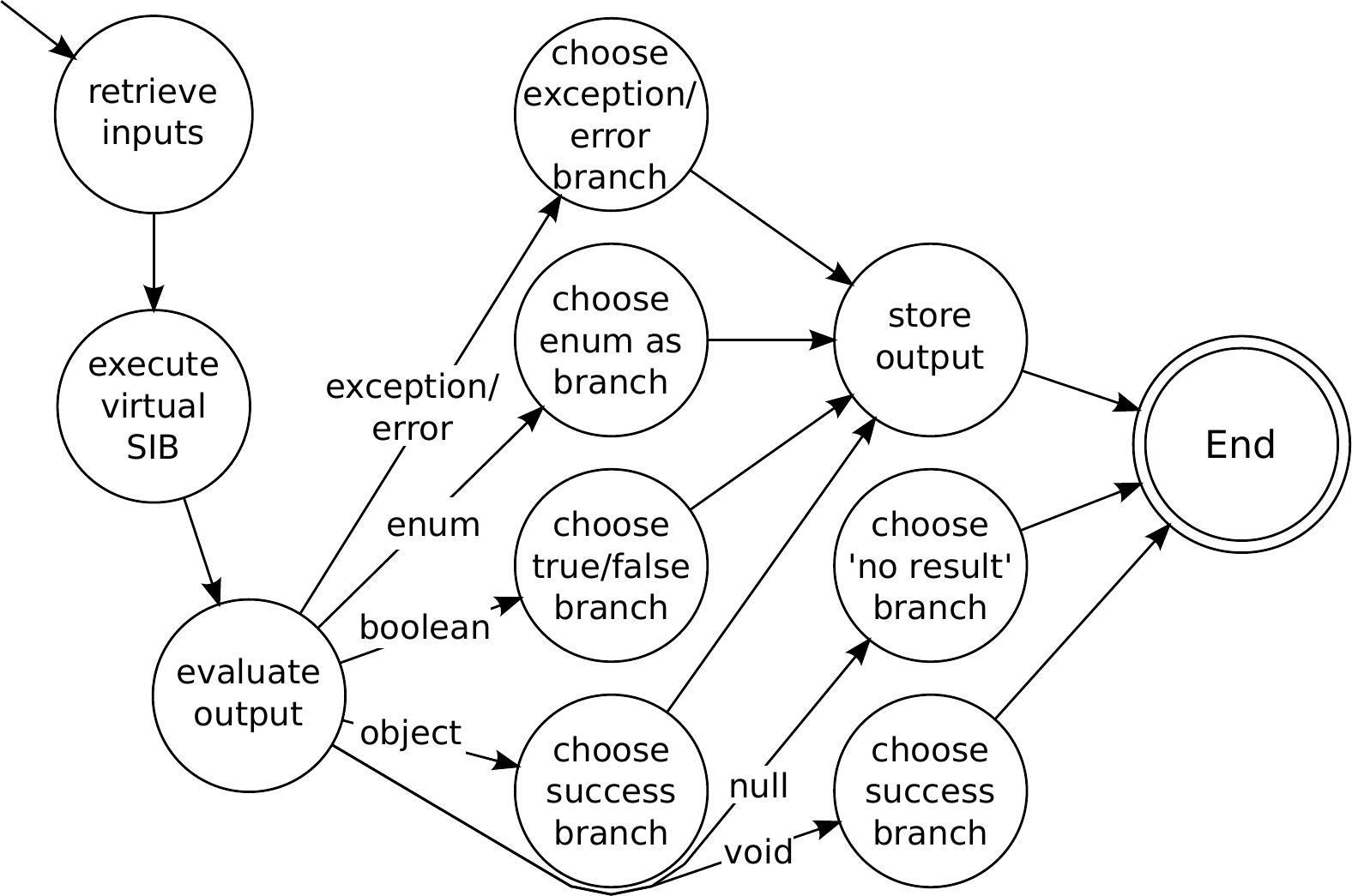}
\caption{Execution of a virtual (dynamic) SIB.}
\label{fig:exec_virtual_sib}
\end{figure}
Fig.~\ref{fig:exec_virtual_sib} shows what happens during the execution of a virtual (dynamic) SIB. We call \newnotion{SIB container} the component that does the pre- and post-processing of the execution. An activity representing a
static method is executed when the control-flow reaches it. Activities corresponding to methods to be executed on a service instance have an input parameter that defines the corresponding context variable. Hence modelers define at modeling
time in a graphical development environment from which context variable the service instance should be
(dynamically) retrieved at runtime. Usually there is a chaining effect: the output parameters of any business activity
may be connected to a context variable, so that at runtime the output of the executed method is stored in context
variables for later use by other activities. Adding back edges even leads to a business activity that executes a method
virtually on different service instances, possibly even of different sub-classes, in the same process. This means that the
process model is able to adapt itself.

As service instances may be reused during an execution and a business activity may execute arbitrary public methods on
it, these services are stateful. In addition, since we exploit the Java type system, we are able to define a business
activity that executes a public method on an interface or superclass. At runtime a context variable may reference an
instance of an arbitrary sub-class of that interface or superclass and may even be exchanged at runtime by the process
itself as mentioned before.

In essence, the dynamic service integration (see right triangle in Fig.~\ref{fig:adapter-dynamic-pattern}) approach is
structured as follows:
\begin{itemize}
  \item Services are provided as methods of a Java
  class or interface, which abstract from technological detail. These are integrated as activities (dynamic SIBs) in
technical models, which are absolutely unaware of the process models.
  \item A fully configured instance of a subclass of the service class or interface is stored to the shared memory of
  a process, denoted by~\newnotion{context}.
  \item The above mentioned instance is provided as input to the corresponding dynamic SIB.
\end{itemize}
The arising technical models are bundled to SIB libraries and published for simple integration into
 high-level, coarse-grained, and domain-specific business process models by an application expert.

We store and retrieve the service instances from the context, considering services as~\newnotion{first class objects}
and therefore introducing a~\newnotion{second-order context} for exchange between the services of a service graph, a
step reminiscent of higher-order functions in functional programming languages~\cite{higherorder}. The dynamic service
integration approach allows for:
\begin{itemize}
 \item on-the-fly dynamic service selection and binding,
 \item virtual service invocation (regarding polymorphism) and therefore static and dynamic variability on the service
  level,
 \item type-safety for service invocation as well as input and output parameters,
 \item separation of concerns between the process model design and the implementation of services,
 \item self-adaption of process models.
\end{itemize}
\subsection{Second-Order Process Models}
\label{sec:2ndorder-pm}
\begin{figure}[t]
\centering
\includegraphics[width=0.4\textwidth]{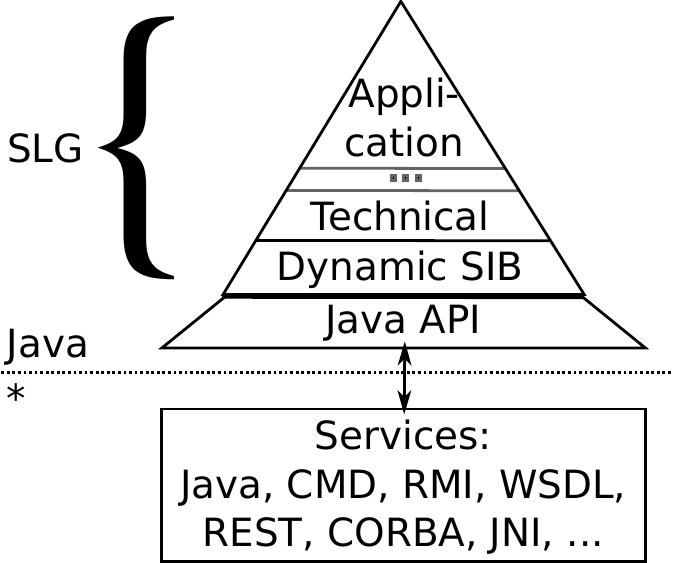}
\caption{The dynamic service integration pattern.}
\label{fig:dynamic-sib-triangle}
\end{figure}
Business activities representing whole sub-models lead to a hierarchical graph structure (cf.
Fig.~\ref{fig:dynamic-sib-triangle}), where coarse-grained SLGs classified as `Application' (level) are
successively refined by more technical SLGs indicated by `Technical', which in turn are based on dynamic SIBs (see
Sec.~\ref{sec:2ndorder-services}). The available type information is used to check whether the use of services and
sub-graphs is sound. This allows us to narrow the semantic gap and to move the technological break behind a Java API,
where only technical experts are involved, enforcing this way the principle ``simple for the many, difficult for the
few''~\cite{1521210}.

Having polymorphic virtual services does not allow one to change the structure of process models.
We therefore recently added support for using process instances as first class objects
in the execution context as well, introducing a notion of second-order process 
modeling~\cite{DBLP:conf/icsob/NeubauerS13}:
context variables referencing process instances may be used as a refinement of business activities, which now
cause the execution of the respective process instance at runtime.

Polymorphic second-order executable process models arise naturally when introducing
\newnotion{interface graphs} (first mentioned in~\cite{DBLP:conf/icsob/NeubauerS13}): they provide the type information 
of
the input and output parameters for derived process models and hide the internal structure of the process models that lay
beneath. In particular, this supports the exchange of the underlying technology, service implementation, or
even of a complete library of service graphs, and hence the change of a process model's behavior at the modeling level
without touching it. Technically this is achieved by equally treating graph and service data-types by exploiting and
enhancing the Java type system.\medskip

\noindent
In the following we exemplify second-order process modeling via our running example. In Fig.~\ref{fig:2ndorder-payment}
we see in the main window a refinement of the process `register to conference' of
Fig.~\ref{fig:static-ocs} that adds a dedicated `get payment process' business activity and shows below it some possible
instantiations of the payment service. OCS users might preconfigure one of them as a default, so that it is retrieved in
the sub-process `get payment process' or they might selected one manually via a combobox or similar on the OCS web page just before they confirm the registration. All these payment services must implement the same input/output
parametrization defined by the interface graph \code{Payment}.

Connecting the second-order
business activity `pay conference fee' to the respective payment service (which may not be determined until runtime) requires four steps:
\begin{enumerate}[(a)]
 \item Add the payment interface graph as a business activity to the SLG via drag and drop from the browser area to the
  graph canvas and name it `pay conference fee'.
 \item Connect the registration information of the output parameters of the business activity `fill registration
  info' to the context variable `registrationInfo'. This guarantees that the registration information will be written to the context variable whenever the corresponding business activity has been executed.
 \item Connect the payment process of the output parameters of the activity `get payment process' to the context
  variable `paymentProcess'.
 \item Connect both registration information and payment process to the inputs of the interface graph activity `pay conference  fee', in order to guarantee that at runtime the payment process will enact the activity `pay conference fee' with the corresponding registration information.
\end{enumerate}
Any of the selected sub-processes is guaranteed to be conform to the interface specifications, since this is already
checked at modeling time.
\section{Higher-Order Modeling}
\label{sec:higher-order}
The XMDD approach allows for complex hierarchical graph structures by means of business activities that represents a complete sub-graph.  We discussed in Sec.~\ref{sec:2ndorder-pm} that we recently added support not only for service
instances, but for whole process instances as first class objects in the context. Now we go one step further to a tamed
variant of higher-order process modeling, which avoids many of the well known problems of full higher-order (cf.
\cite{Langmaack73,Schmidt96}) as it is based on a very strong notion of type correctness (cf. \cite{Schmidt94}).

Sub-process instances may be used as input and output parameters of an SLG and also be stored in the execution context
of every hierarchy level. The actual parameters of a graph type are process instances with its own runtime execution
context, which is again second-order. Hence we slice our higher-order models into second-order contexts, which are
stacked implicitly. This results in a `tamed' notion of higher-order which allows us to type every parameter and 
context
variable and to use the Java type system enhanced for flat, simplified types for process models at modeling time.

Hence besides selection of existing process models implementing a given interface, a process instance may be provided
via an arbitrary retrieval method, via a `retrieval sub-process' activity or via an input parameter of the overlying
process at runtime. This ranges from ad-hoc process changes~\cite{AristaFlow,Lampre2012} over discovery
mechanisms~\cite{swsmiAamics} to process-model synthesis~\cite{LaNaMS2010}. In case such retrieval sub-processes are
themselves passed as parameters, we arrive at third-order processes, or even at arbitrary higher-order processes by
continuing this line of process construction.

In the following subsections we will elaborate on the concepts behind higher-order modeling, illustrate its outreach,
and its realization by showing how the execution of higher-order models is realized for our running example. In
the higher-order example we
\begin{itemize}
 \item replace the sub-process for the activity `validate proceedings' with a synthesized variant of the loosely
specified process model of Fig.~\ref{fig:ho-synthesis},
 \item which itself uses an ad hoc modeled variant of the `validate payment' process model in the activity `registered',
 \item which uses different payment service SLGs for the activity `did at least one author pay?'.
\end{itemize}
\subsection{Concepts of Higher-Order Modeling}
\label{sec:idea}
The main technical idea of second-order process modeling is to allow virtual process model calls in form of business
activities representing sub-models, introducing polymorphic executable process models. This is realized by considering
process model instances, denoted by \newnotion{process instance}, as~\newnotion{first class objects} and therefore
introducing a~\newnotion{second-order context} for exchange between the sub-graphs of a service graph.

For higher-order modeling, we go one step further towards higher-order functions known from functional programming, as
a process instance may have a context with further process instances and so forth. This is realized via a local context
for every hierarchy level, that is implicitly stacked. Hence, our context is still second-order. Furthermore we offer
activities for the creation of new process instances denoted by \newnotion{constructor activity}.  They may be
parameterized with initial values like a constructor in programming languages.

Besides the advantages of the above mentioned dynamic SIB pattern, we now lift polymorphism and loose coupling of services
to all hierarchy levels, supporting all-pervasive high reusability and variability not only at modeling-time but at
runtime, too. This even leads to self-adaptive process models. At the same time type-safety is preserved on the process
model level by employing and enhancing the type system of Java. Type-safety is not limited to primitive types. We 
support
domain-specific complex types like \code{User} and \code{Paper}, but also the type of a graph (interface and service
graphs) that are not represented by Java types but via a meta model for SLGs.

Service activities (dynamic SIBs) are represented by Java methods and their signature is used as information on input
and output behavior. For process models, this information is modeled in so called \newnotion{service graphs} and
\newnotion{interface graphs}. The development environment ensures that derived service graphs fullfil the requirements
of their interface graph. The graphs as well as Java methods have fully defined interface descriptions and hide
their internals. This supports loose coupling and reusability as well as the organization of process models in process libraries and is technically achieved by:
\begin{itemize}
 \item equally treating services (dynamic SIBs) and SLGs as (business) activities,
 \item constraining the context of processes to local access from and to the current hierarchy level,
 \item canalizing communication between hierarchy levels via the input/output behavior of an SLG,
 \item storing and retrieving the instance of a process model to/from the context, in order to be able to exchange the
  process implementation of an interface activity, and
 \item documenting service and interface graphs and their input and output parameters in a similar fashion as, e.g.
  Javadoc for Java.
\end{itemize}
Hence, interface graphs, their process model implementation (service graphs), and Java methods (services) equally are
treated as activities in a process model. The former introduce hierarchy, since they represent the execution of the
respective sub-graph. The latter are represented by arbitrary Java methods. The documentation is presented in the GUI
of the development environment jABC as tooltips in order to ease communication between the modeler of different
hierarchy and therefore abstraction levels.

The interface and service graphs are organized in SLG libraries and are published to an~\newnotion{application expert}
for use without any knowledge about the details of the underlying structure. This may even be used for communication
of requirements, as an application expert may, e.g., model and document a dummy service or interface graph, that is
then implemented by someone else. The realization of the hierarchy levels can be done separately, by different
responsible people or groups. Every graph (interface and service graphs) may be equipped with an icon (e.g. a raster or
vector graphic) which is used for its representation as it is used as a business activity. This eases the understanding
of a model with a lot of sub-processes and serves as a recognition feature that appeared to be very effective in
cooperation with industrial partners.

Our new approach lifts the variability already gained via polymorphic services to the modeling-level, i.e., it
is only a matter of design on which level of abstraction polymorphism is introduced. This has different
dimensions on all hierarchy levels, e.g.:
\begin{itemize}
  \item Variants of a process step in a high-level process can be implemented differently introducing product-lines,
e.g. having an interface graph for approval or authorization (of documents)  that may be implemented as a four-eye
principle, via peer reviewing, or via counter-sign. These may be reused in (or shared between) different applications.
  \item a high-level process may focus on the structure as the complete implementation details, e.g. whether user
input comes from a terminal, a GUI, a web page, or a mobile device, are hidden from the application expert completely.
  \item the level of abstraction can be changed by choosing implementations that operate on different layers of a
  multi-layered application like, e.g., the presentation layer and the business logic layer.
\end{itemize}
A business activity representing a service or interface graph SIB is denoted by \emph{interface graph SIB} resp.
\emph{service graph SIB}, or \emph{graph SIB} if any of both is meant. In general, their interface description consists
of a list of inputs, the origin of the process instance, a list of outputs for each outgoing branch, and a reference to
its interface or service graph. Process instances as well as outputs are read from resp. written to the context.

The corresponding values are represented by context variables at runtime. A process instance may be initialized and put
to the context in the graph itself (via constructor SIBs) or before the process has been started. Input parameters may
be provided statically for primitive types as well as a selection of Java types supported by the jABC, e.g., strings,
enumerations, and file handles. That means that a modeler can design new SLGs and decide which input values are
statically entered and which are retrieved from shared resources on-the-fly at modeling-time.
\subsection{The Power of Higher-Order Modeling}
In this section we show how our approach can be combined with automatic synthesis technology in order to dynamically
include runtime generated processes into the process execution. This captures the realization of ad hoc modeling, which
is the simple case where the synthesis is performed manually.
\subsubsection{Higher-Order Modeling with Synthesis}
\label{sec:ho-synthesis}
\begin{figure}[t]
\centering
\includegraphics[width=.5\textwidth]{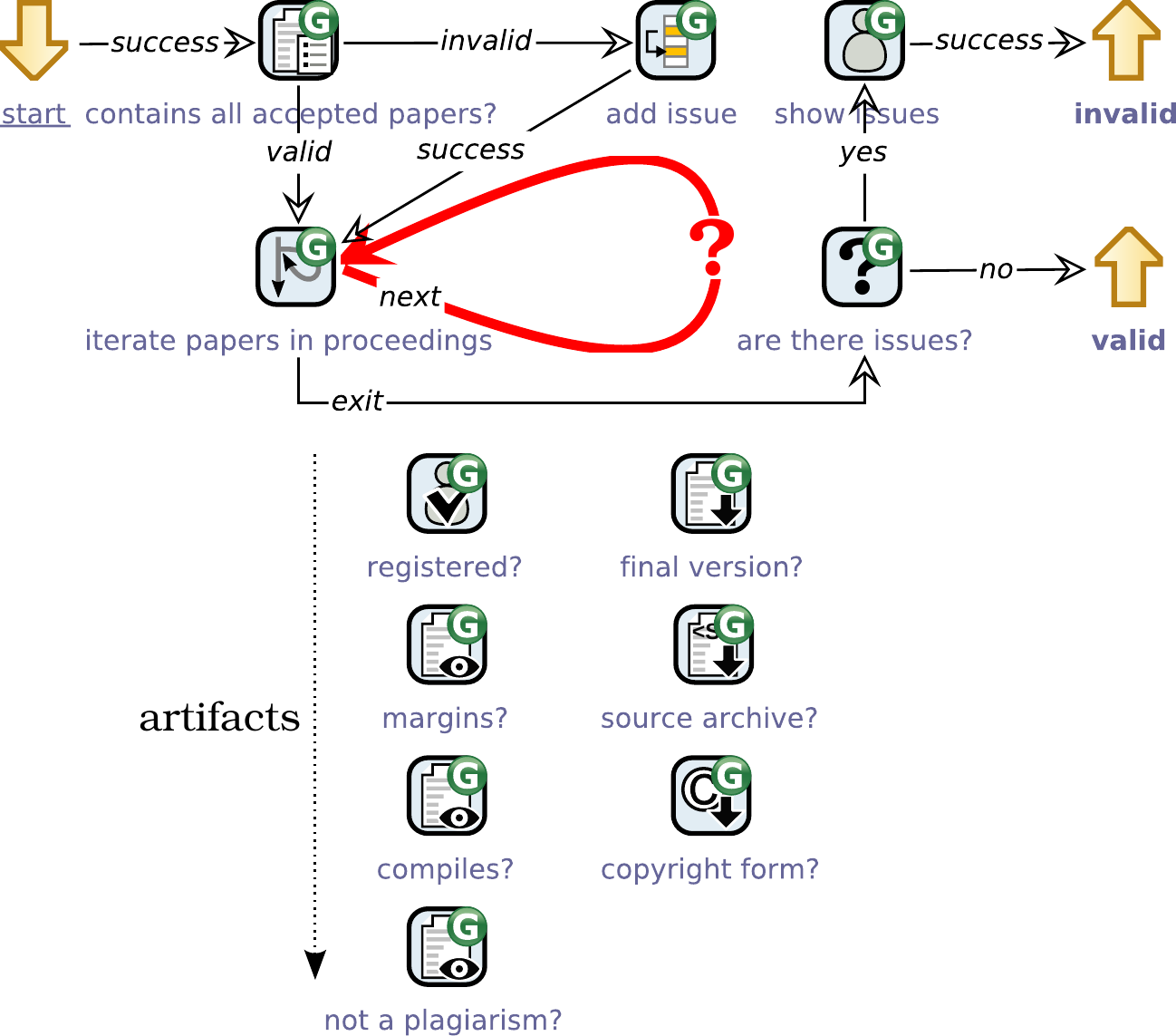}
\caption{SLG with underspecified  `next' branch of the activity `iterate papers in proceedings'.}
\label{fig:ho-synthesis}
\end{figure}
We introduce runtime variability by keeping process instances in a type-safe context and execute them when a business
activity is reached that references the corresponding context variable. The business activity may reference an
interface graph which defines the type-safe input/output parametrization. An instance of any process model
implementing this interface may be used. In jABC one can annotate interface graphs and concrete
SLGs with further constraints like, e.g., model checking formulas expressing a \newnotion{Service Level Agreement} (SLA).
This information can be used to aid the synthesis-based completion of loosely specified process models\cite{LaNaMS2010}.
The PROPHETS plugin of the jABC offers process synthesis along  the `loose programming' paradigm~\cite{NaLaSt2012}.

\begin{figure}[t]
\centering
\includegraphics[width=.6\textwidth]{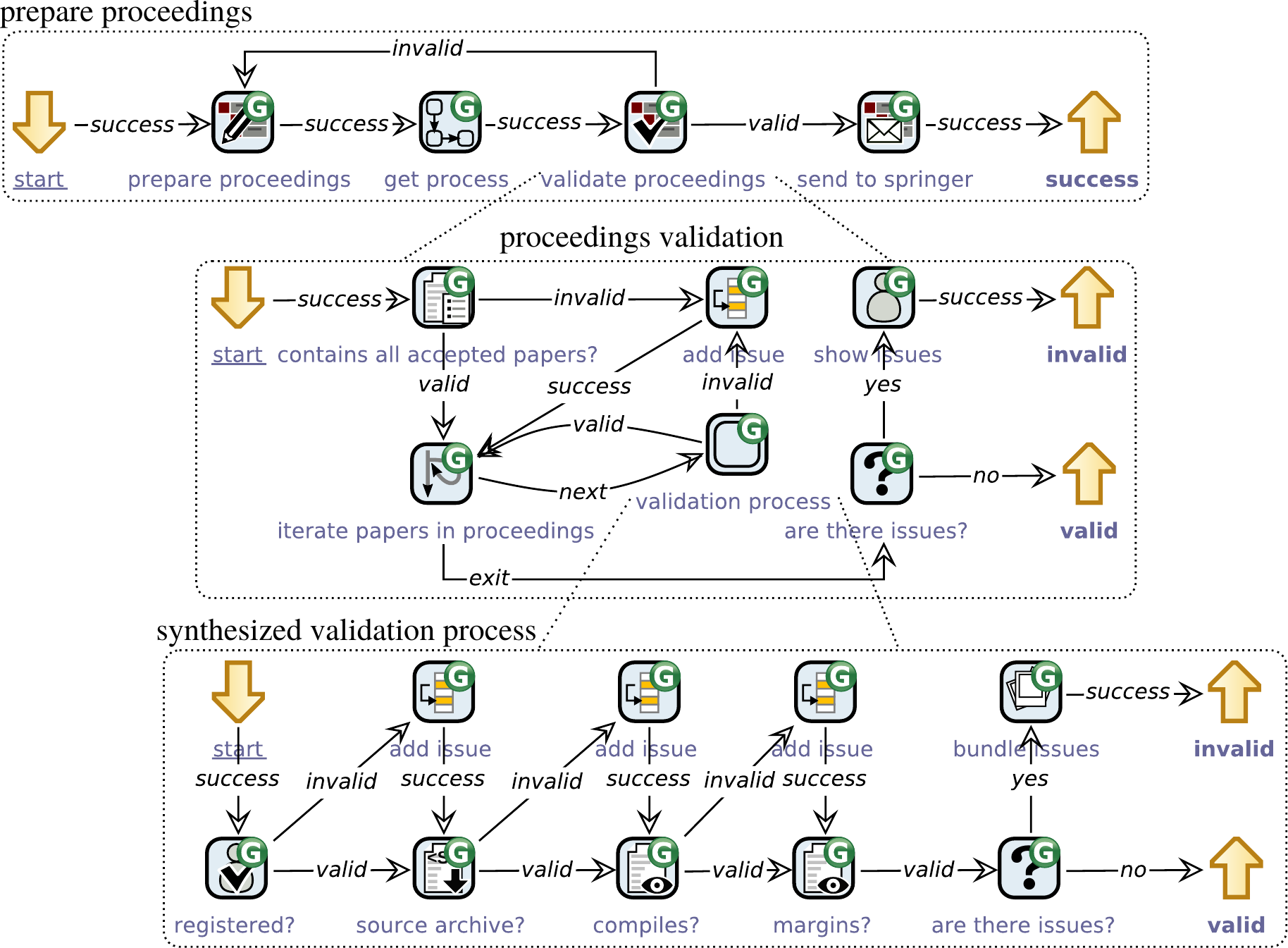}
\caption{A fully specified synthesized variant of the `simple proceedings validation' process in
Fig.~\ref{fig:ho-synthesis}. The loosely specified branch `next' of the activity `iterate papers in proceedings' now
executes the synthesized sub-process model `synthesized validation process'.}
\label{fig:ho-synth-exec}
\end{figure}

Fig.~\ref{fig:ho-synthesis} shows a (loosely specified) version of the process `simple proceedings validation' (cf.
Fig.~\ref{fig:static-ocs}), where the `next' branch of the activity `iterate papers in proceedings' is
unspecified (as indicated by the label `?'). With our synthesis technology we are able to dynamically replace at runtime the unspecified
`next' edge of `iterate papers in proceedings' with the online synthesized `synthesized validation process'
as shown in Fig.~\ref{fig:ho-synth-exec}.
The required synthesis algorithm works by orchestrating the following business activities:
\begin{description}
 \item[registered?] validates whether the user is registered to the conference.
 \item[margins?] tests whether the margins of the paper document satisfy the required style.
 \item[compiles?] checks whether the uploaded source files compile to a valid PDF document and store the result in
  the context.
 \item[not a plagiarism?] uses an external plagiarism checker for validating the uploaded PDF document.
 \item[final version?] validates whether a final document has been uploaded and stores it in the context.
 \item[source archive?] checks whether a source archive has been uploaded.
 \item[copyright form?] tests whether a copyright form has been uploaded.
\end{description}
while exploiting the following extra knowledge formulated in temporal logics:
\begin{align}
\label{eq:data-flow1}
\neg \textrm{`not a plagiarism?'}~ \textrm{WU}~ (\textrm{`compiles?'}~ \vee~ \textrm{`finalVersion?'})\\
\neg \textrm{`margins?'}~ \textrm{WU}~ (\textrm{`compiles?'}~ \vee~ \textrm{`finalVersion?'})\\
\label{eq:data-flow3}
\neg \textrm{`compiles?'}~ \textrm{WU}~ \textrm{`sourceArchive?'}\\
\label{eq:real-constraints1}
\textrm{F}(\textrm{`margins?'})
\end{align}
The formulas~(\ref{eq:data-flow1})-(\ref{eq:data-flow3}) describe the data dependencies between the activities. The
check for plagiarism and for correct margins need a PDF document, which will be available in the context if either a
final version has been uploaded or the sources are compiled successfully. Formula~(\ref{eq:real-constraints1}) demands
that a margins check has to be in the synthesized solution.

Our synthesis plugin PROPHETS exploits data-flow information computed via the {\em dataflow analysis as model
checking} paradigm \cite{Steffe1991, Steffe1993, Schmidt98, SchSte1998}, which automatically takes care of the
formulas~(\ref{eq:data-flow1})-(\ref{eq:data-flow3}) as part of our correctness by construction approach~\cite{DBLP:conf/fmco/LamprechtMSS11}.
\subsubsection{Ad-Hoc Modeling}
\begin{figure}[t]
\centering
\includegraphics[width=.75\textwidth]{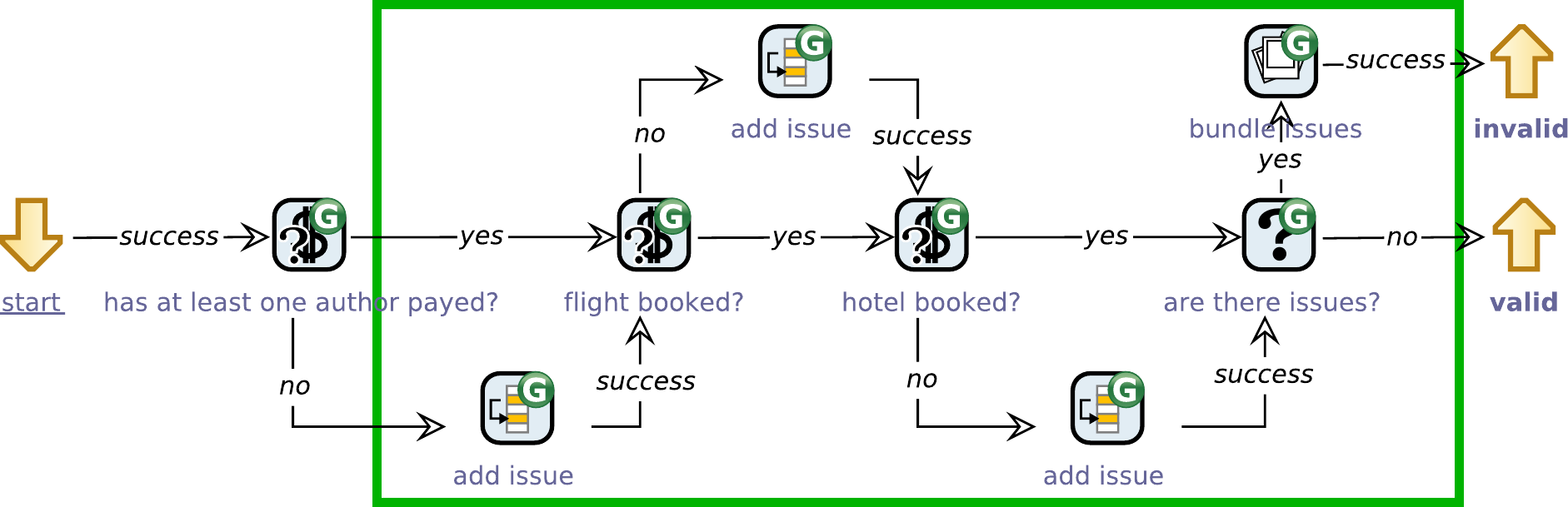}
\caption{An ad-hoc changed version of the process model `validate payment' in Fig.~\ref{fig:static-ocs} that additionally tests
whether at least one author of the paper has booked a flight and a hotel for the conference.}
\label{fig:ho-ad-hoc}
\end{figure}
In this section we illustrate how  the sub-process
for the activity `registered?' in the synthesized validation process of Sec.~\ref{sec:ho-synthesis} can be replaced by an ad hoc changed version of the process model `validate
payment' in Fig.~\ref{fig:static-ocs} that adds additional tests. Fig.~\ref{fig:ho-ad-hoc} shows
the ad hoc adapted process model. The modified part is emphasized via a rectangle. Besides checking
whether at least one author has payed the conference fee, it checks also the booking of a flight and a hotel. This reveals authors who pay the conference fee, but might not show up at the conference. The process
model still conforms to the interface graph for process validations and therefore can be seamlessly used in the process
model `synthesized validation process' in Fig~\ref{fig:ho-synth-exec}. This combined example shows that we can easily
support runtime higher-order process modeling because every hierarchy-level is  on the one hand loosely coupled, but on the
other hand correct by construction, thanks to type safety and the fact that the SLAs are controlled by model checking~\cite{MuScSt1999} in our framework.

\subsection{Realization}
\label{sec:realization}
\begin{figure}[t]
\centering
\includegraphics[width=.6\textwidth]{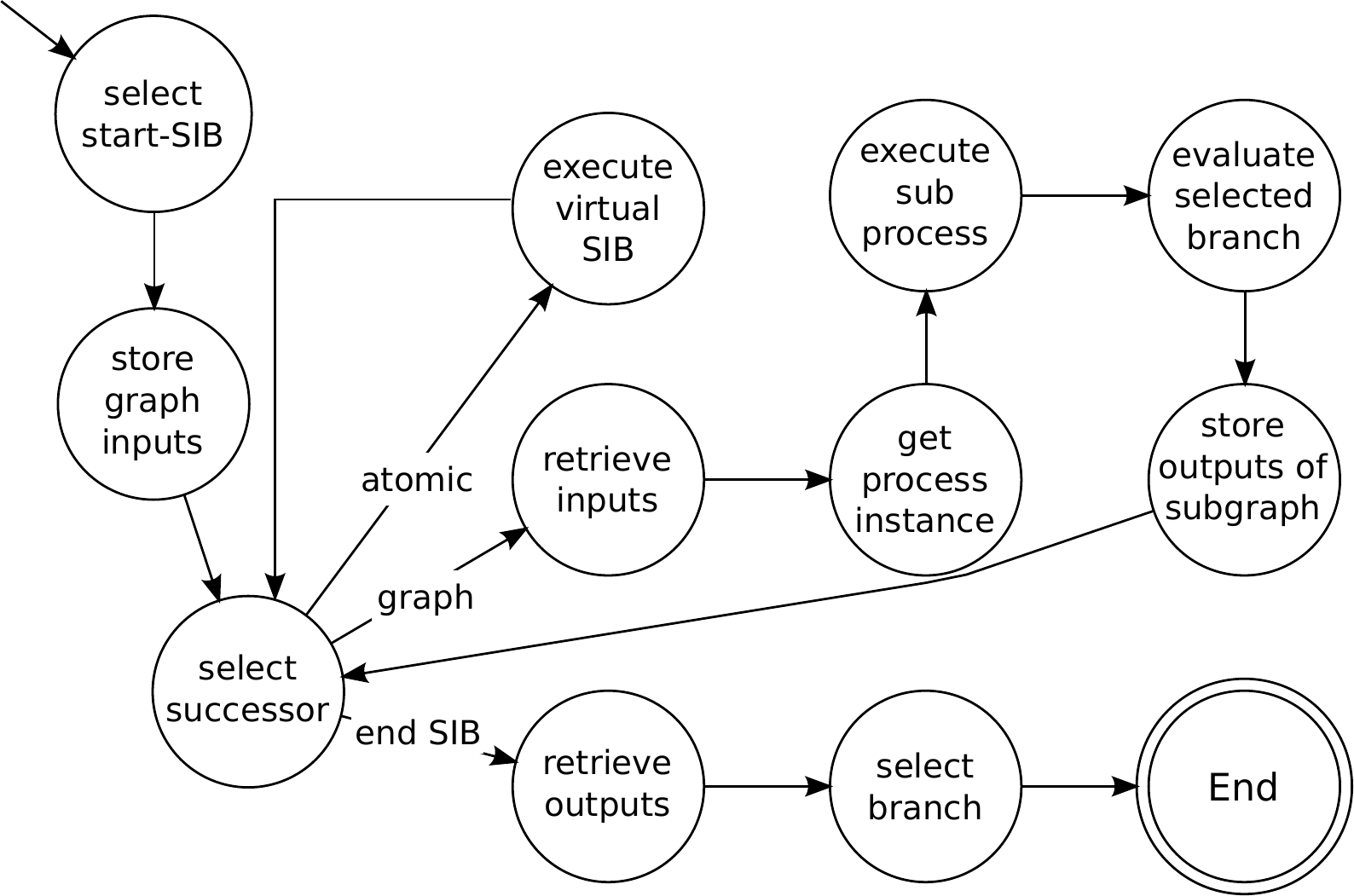}
\caption{Steps during execution of a higher-order SLG.}
\label{fig:execution}
\end{figure}
In the following we will detail how the higher-order modeling works step by step by means of a running example
(refer Sec.~\ref{sec:ocs}).
\subsubsection{Execution of a SLG}
The procedure of a service graph invocation is as follows (cf. Fig.~\ref{fig:execution}). The interpreter
\begin{itemize}
 \item retrieves the actual type-safe input parameters of the graph defined in its start-SIB and stores them into the
context,
 \item selects the successor activity of the start-SIB,
 \item in case it is an atomic (virtual) SIB, it is executed as illustrated in Fig.~\ref{fig:exec_virtual_sib}
 \item in case it is a graph SIB the interpreter:
 \begin{itemize}
  \item provides the required input parameters as defined in the graph SIB, i.e., either from the context or
  static values provided during modeling,
  \item retrieves the corresponding process instance from the type-safe execution context,
  \item executes the sub-graph with the given process instance,
  \item as the execution returns it evaluates which outgoing branch has been determined and puts the regarding output parameters
  into the context as defined by the graph SIB, and
  \item follows the chosen branch, in order to find the successor SIB.
 \end{itemize}
 \item in case of an end-SIB the interpreter:
 \begin{itemize}
  \item retrieves the defined output parameters from the local execution context,
  \item selects the output branch, and
  \item returns to the hierarchy-level above or in case of a root process it terminates.
 \end{itemize}
\end{itemize}
The access to the context is transparent for the services. The interpreter evaluates the context and static inputs to a
dynamic SIB and passes them to the service call. Furthermore the type-safety regarding the process instance is ensured
by enhancing the type system of Java via a data-structure that may represent a Java type or a graph type, in order to
distinguish SLGs as dedicated types although there is no corresponding Java type available. This is true for both
service and interface graphs as well as the inheritance structure between them. For a service graph SIB, the declaration
of a context variable may be omitted. In this case a new process instance is created as needed.

Currently we use an interpreter to execute our processes. This is in line with the common practice of having interpreting business process engines, e.g., for BPMN. Nonetheless we use the type information in our
enhanced type data-structure to check type consistency statically, therefore type-safety is assured already at
modeling-time. This is true for the process and service instances as well as for every output and input of SLGs
and services.
\subsubsection{Execution of a Graph SIB}
\begin{figure}[t]
\centering
\includegraphics[width=\textwidth]{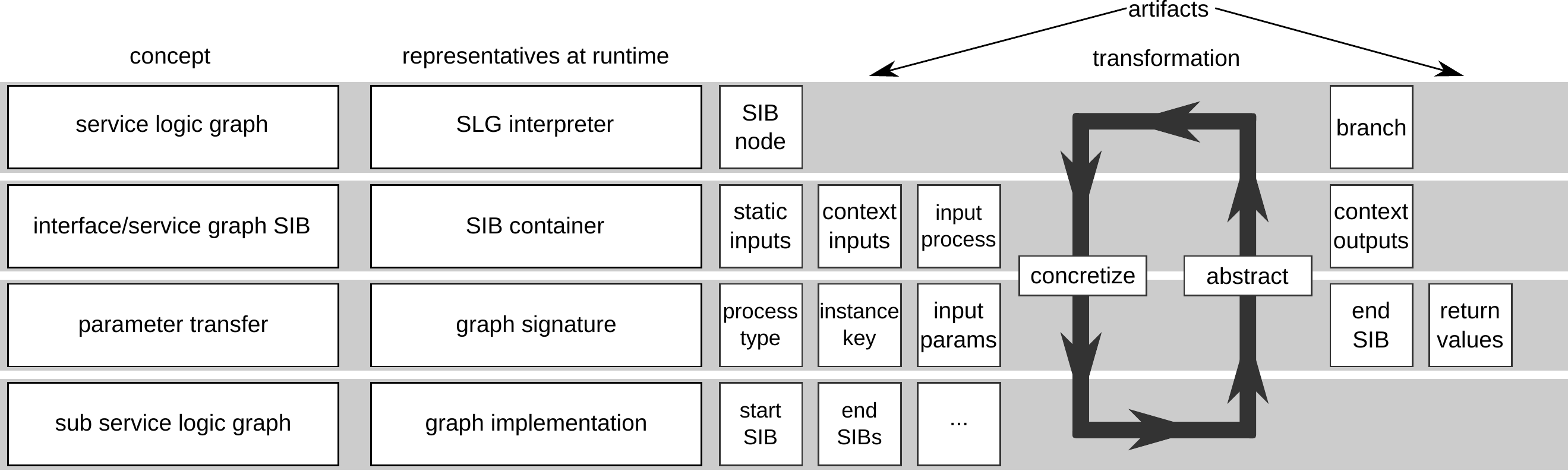}
\caption{\label{fig:delegation}Layer-model of the dynamic SIB pattern execution.}
\end{figure}

Fig.~\ref{fig:delegation} illustrates the layers and artifacts involved in the interpretation of a SLG along the circular flow of one graph SIB execution. The operational flow follows two directions: it first~\newnotion{concretizes} all artefacts from the process model level to sub-process execution, then it~\newnotion{abstracts} to the process model level again, in order to find and execute the successor SIB.

Concretization starts at the highest level, where the service logic graph is executed by an interpreter.
The SLG interpreter selects the current SIB node and delegates its concrete invocation to the SIB container,
responsible for the context access. Here, the static inputs and inputs from the context as well as the process instance
itself are read in as defined in the configuration of the SIB and provided to the sub-process. In particular,
process instances may be retrieved from the context. All this information is handed over to the parameter transfer
layer. A process model is defined via its graph-type, its input parameters, its branches, and the output
parameters of each branch, which altogether form its signature. As process instances can be any implementation of an
interface graph, they need to be invoked virtually.

After the sub-process has returned, the abstraction phase starts. At first the end SIB and the related branch as well
as the corresponding output values are retrieved. Then the value is put to the context as defined in the SIB
configuration. Finally, the corresponding branch is evaluated and the successor SIB is selected in the SLG.
\subsubsection{Application to the Running Example}
The `prepare proceedings' process model in Fig.~\ref{fig:ho-synth-exec} has an activity `get process' that
synthesizes the loosely specified branch in the process model `proceedings validation' (cf.
Fig.~\ref{fig:ho-synthesis}) and puts the process instance into a context variable. The following SIB is an interface
graph SIB with the graph-type \code{ProceedingsValidation}, which our process model `proceedings validation'
and all synthesized variants implement.
\subsubsection{Putting Together}
\begin{figure}[t]
\centering
\includegraphics[width=\textwidth]{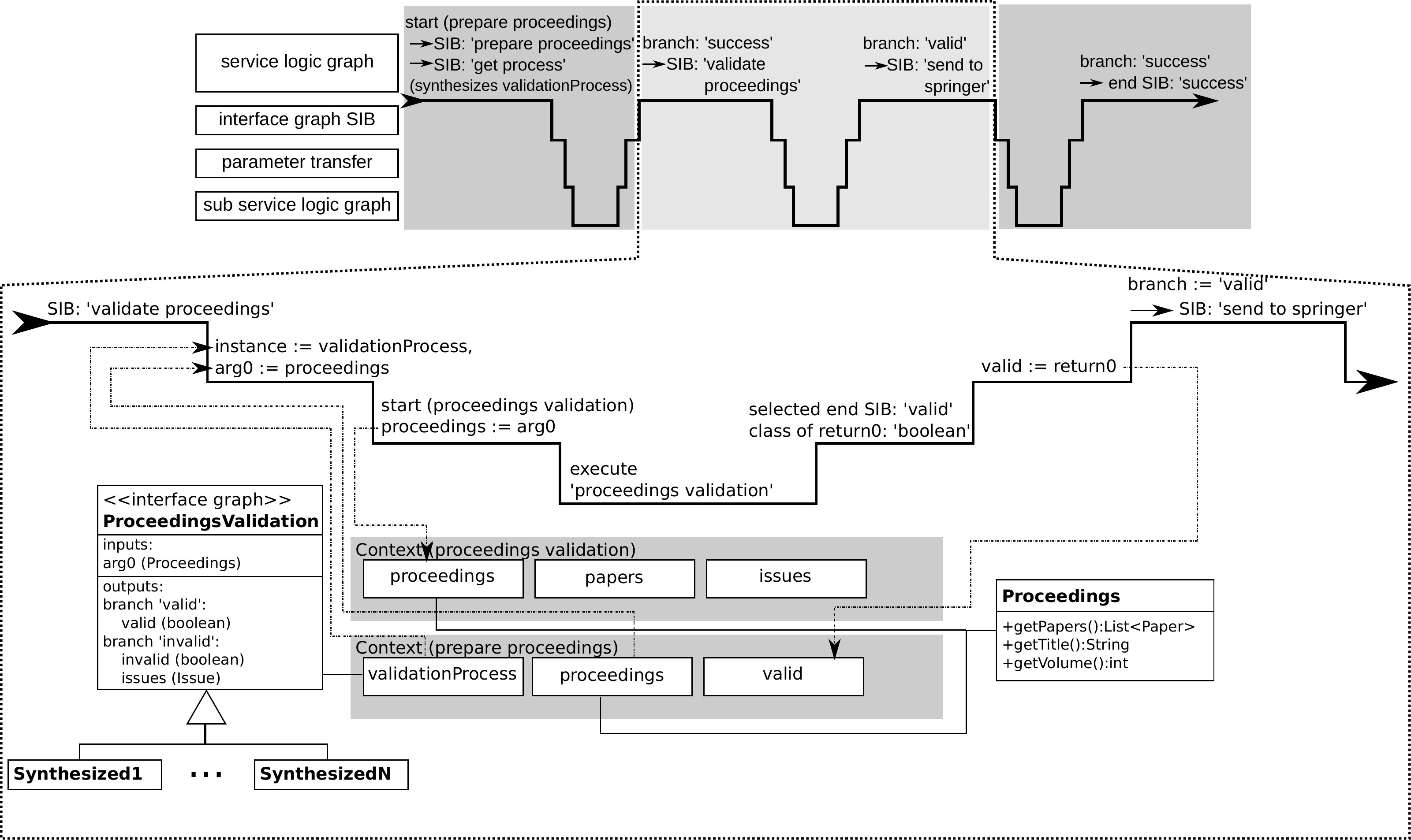}
\caption{Run through the different layers for a specific graph SIB execution.}
\label{fig:sawtooth}
\end{figure}

Fig.~\ref{fig:sawtooth} shows an excerpt of the concrete execution of the example SLG in Fig.~\ref{fig:ho-synth-exec}. The top of Fig.~\ref{fig:sawtooth} depicts a coarse overview of the execution exhibiting the concretization and
abstraction process for each invocation of a higher-order graph SIB. The excerpt begins with the start SIB of the `prepare proceedings' process model, followed by the business activity
`prepare proceedings', and `get process'. Both executions are successful, therefore the control-flow follows the
`success branches'. The next SIB is the interface graph SIB `validate proceedings', which we will discuss in more
detail.

Fig.~\ref{fig:sawtooth} (bottom) depicts in the dotted polygon the call of the graph SIB `validate proceedings'.
The thick line shaped like a sawtooth represents the execution flow through the different layers
of the above mentioned concretization and abstraction process. The implicitly stacked contexts of the two hierarchy levels are illustrated at the bottom of Fig.~\ref{fig:sawtooth} in grey boxes containing the context
variables. The dotted directed edges connecting context variables describe the data-flow. The edge
starting in~\code{validationProcess} and pointing at~\code{instance := validationProcess} indicates that the current
value of the context variable is written to the input parameter~\code{instance} of the activity `validate proceedings'.
Each context variable is associated to a Java or graph type, which can be a concrete class like~\code{Proceedings}, an
interface, a service graph (like \code{Synthesized1-N}), or an interface graph (like \code{ProceedingsValidation}). A
context variable can be used as input, instance, or output of a process model.

The invocation of the graph SIB starts on the service logic graph layer. Depending on the selected branch of the
predecessor, the SLG interpreter selects the activity to execute, in our case `validate proceedings'. On the
next layer, the SIB container reads the process instance from the context variable \code{validationProcess}, which is a
sub-type of the interface graph \code{ProceedingsValidation}. In addition the input `arg0' is read from the
context variable `proceedings', which is an instance of the domain-specific Java type \code{Proceedings}. The
parameter transfer layer evaluates the graph signature. It defines that the service instance must be a sub-type of
\code{ProceedingsValidation}, and the graph interface defines one argument of type \code{Proceedings}. The SLG
interpreter executes the respective sub-process on the given process instance and passed arguments with its own
context. The start SIB of the sub-process stores the input parameter `arg0' to the context variable `proceedings' in the
context of the sub-process. At this point the concretization process is completed and the execution waits until the sub
process returns.

As the sub-process returns, the end-SIB is evaluated, the related branch is chosen, then the return values are
analized. There is only one output parameter of type \code{boolean} for the selected output branch denoted `valid'.
Next, the return value is put to the output context variable analogously named `valid'. On the service logic
graph layer, the corresponding successor SIB~\code{send to springer} is invoked. Now, the abstraction phase is finished
and the following SIB execution starts as outlined in the above mentioned coarse-grained flow description.
\section{Conclusion and Perspectives}
\label{sec:concl}
We have presented a comprehensive, graphical binding and execution framework for (business) process models tailored to
support the integration of 1) ad hoc (graphically) modeled processes, 2) third party services discovered in the (Inter)
net, and 3) (dynamically) synthesized process chains solving situation-specific tasks not only at design time, but also
at runtime. Key to our approach is the introduction of stacked type-safe second-order execution contexts allowing for
(tamed) higher-order process modeling which, together with our underlying strict service-oriented notion of 
abstraction, is tailored also to be used by application experts, because it is easy to show to end-users only the 
portions of the hierarchy that remain at the application level, shielding these users from the technical level. We even 
forbid recursion on an inter-process level in order to further simplify the semantics in~\cite{NeuSte2013}.

The experience with our new framework, which allows users to select, modify, construct and then pass (component)
processes during process execution as if they were data, is very promising.
 In particular, the most advanced feature of our new framework, which allows one to combine
online synthesis with a corresponding integration of the synthesized process into the running application may have a strong impact on the current modeling style. In fact, it has the potential to allow even application experts
to loosely specify their demands in a graphical fashion both at design time (cf. \cite{NaLaSt2012}) and at runtime, with
the guarantee that the frameworks' underlying higher-order type discipline together with its corresponding automatic
integration and binding process guarantee executability.

More concretely, we have shown how executable higher-order process models arise naturally when lifting process contexts to
(stacked) second-order. We also show how this approach
supports a clear separation of concerns, when combined with the eXtreme Model-Driven Design paradigm (XMDD):
\begin{itemize}
  \item it separates the implementation of services from domain-modeling concerns by means of low-level service graphs that bind ``the right'' services
  dynamically and polymorphic, and
  \item it makes these low-level SLGs accessible beyond single projects in order to use them as coarse grained service definitions  in  a wealth of application-level service graphs.
\end{itemize}
The service binding is done directly within the process models, so we have control regarding the access to the shared resources of a process. This allows us to perform static type-checking in order to avoid problems at runtime. Having the
complete Java method signature information at hand, this approach captures also the type-safety of our second-order 
service
binding approach, which reaches far beyond the state of the art of today's business process management frameworks.
The impact of our approach has been illustrated in the context of modeling the variant rich Proceedings Preparation
Service (PPS) of Springer's Online Conference Service OCS.

Our case study with the OCS clearly illustrates the power of our higher-order approach to organize software product lines and to
achieve variability at the modeling level. We are therefore currently investigating the deeper impact of our higher-order
approach as a unified backbone for product-line management, variability modeling, and self-adaptation technology.

\bibliographystyle{eptcs}
\bibliography{references}   

\end{document}